%% file: main.tex
\documentclass[12pt]{article}
\usepackage{geometry}
\usepackage{graphicx} 
\usepackage{caption}
\usepackage{subcaption}
\usepackage{stfloats}
\usepackage{amsmath,amsthm,amssymb,amsfonts,mathtools,bbold}
\usepackage[colorlinks=true
  ,urlcolor=blue
  ,anchorcolor=blue
  ,citecolor=blue
  ,filecolor=blue
  ,linkcolor=blue
  ,menucolor=blue
  ,linktocpage=true
  ,pdfproducer=medialab
  ,pdfa=true
]{hyperref}
\usepackage{xcolor}
\usepackage{enumerate}
\usepackage{cite}
\usepackage{color}
\usepackage{tikz}

\newcommand{\hs}{\mathcal{H}} 
\newcommand{\swap}{\mathcal{S}} 
\DeclareMathOperator{\tr}{tr} 
\newcommand{\B}{\mathcal{B}} 
\newcommand{\id}{\mathbb{1}} 
\DeclareMathOperator{\baby}{baby} 
\DeclareMathOperator{\nb}{nb} 

\newcommand\numberthis{\addtocounter{equation}{1}\tag{\theequation}}

\begin{document}

\begin{titlepage}
{\ }
\vskip 1in

\begin{center}
{\LARGE On tests for baby universes in AdS/CFT}
\vskip 0.5in 
{\Large Kenneth Higginbotham\footnote{kenneth.higginbotham@colorado.edu}}
\vskip 0.2in 
{\it Department of Physics \\
390 UCB \\ University of Colorado \\ Boulder, CO 80309, USA}
\end{center}
\vskip 0.5in

\begin{abstract}\noindent
To address a puzzle by Antonini and Rath -- where a single CFT state has two bulk duals, one with a baby universe and one without -- Engelhardt and Gesteau recently devised a test for baby universes in AdS/CFT. Using the extrapolate dictionary, they showed that the boundary dual of a bulk swap test favored bulk spacetimes without a baby universe, providing evidence against their semiclassical validity. However, recent work suggests that holographic maps should post-select on such closed universes, and we argue that this is consistent with the extrapolate dictionary. We therefore construct a new holographic map for bulk states with baby universes and use this to show that the swap test cannot distinguish between Antonini and Rath's two candidate bulk duals. This not only allows for a valid semiclassical description of the baby universe, but also enables the application of recent techniques for including observers in holographic maps.
\end{abstract}

\end{titlepage}

\section{Introduction}

Baby (or \textit{closed}) universes pose a challenge to our understanding of holography. Although they are well-described semiclassically, a variety of evidence suggests that the fundamental description of a baby universe is trivially given by a one-dimensional Hilbert space 
\cite{maldacena_Wormholes_2004,
almheiri_Page_2020,
penington_Replica_2020,
marolf_Transcending_2020,
mcnamara_Baby_2020,
usatyuk_Closed_2024,
usatyuk_Closed_2025}. This tension makes baby universes a natural testing ground for further development of holographic theories. For instance, recent proposals for incorporating bulk observers into gravitational path integrals and holographic maps \cite{harlow_Quantum_2025, abdalla_Gravitational_2025, akers_Observers_2025, chen_Observers_2025} have been motivated by the search for a non-trivial fundamental description of baby universes.

A sharp version of this problem was articulated by Antonini and Rath (AR) \cite{antonini_Holographic_2025} in the AdS/CFT correspondence \cite{maldacena_Large_1999,gubser_Gauge_1998,witten_Sitter_1998}. AR identified a CFT state with two equally valid bulk duals: one with a baby universe, and one without. This apparent non-uniqueness challenges the standard view of holography as a one-to-one correspondence, and AR proposed two possible resolutions to this puzzle.\footnote{AR also proposed ensemble averaging as a third way to resolve the puzzle. However, they noted that a concrete understanding of ensemble averaging only exists in 2-dimensional JT gravity \cite{jackiw_Lower_1985,teitelboim_Gravitation_1983,saad_JT_2019}. It is unclear whether averaging can solve the puzzle in higher dimensions.} If the bulk dual is genuinely non-unique, then additional data must supplement the CFT state to recover a unique bulk dual; AR dubbed this hypothesis ``beyond AdS/CFT.'' Alternatively, if we insist that the holographic dictionary is uniquely defined, then AR suggested that we are forced to sacrifice the validity of semiclassical baby universes.

Recent work by Engelhardt and Gesteau (EG) favors the latter hypothesis \cite{engelhardt_Further_2025}. EG studied a bulk swap operator (or ``swaperator''\footnote{We thank Gracemarie Bueller for suggesting this term.}) $\swap$ designed to detect the presence of a baby universe in the bulk state. To construct a boundary operator $\swap_\partial$ dual to $\swap$, EG relied only on the extrapolate dictionary, which was assumed to unambiguously define an isometric holographic map $V$. The expectation values of $\swap$ and its boundary dual were then found to satisfy:
\begin{equation} \label{eq:exp_vals}
    \langle\swap_\partial\rangle_{\text{CFT}} = \langle\swap\rangle_{\text{no-baby}} \gg \langle\swap\rangle_{\text{baby}}.
\end{equation}
Under the assumption of an unambiguously defined $V$, the agreement between $\langle\swap_\partial\rangle$ in the CFT and $\langle\swap\rangle$ in the ``no-baby state'' suggests that AdS/CFT prefers the bulk state without a baby universe.

In this work, we propose that the extrapolate dictionary need not uniquely define the AdS/CFT dictionary. A growing body of work suggests that holographic maps should act non-isometrically on closed universes by post-selecting on a single state 
\cite{antonini_Cosmology_2023,
akers_Black_2024,
harlow_Quantum_2025,
abdalla_Gravitational_2025,
akers_Observers_2025}. We interpret this as consistent with the extrapolate dictionary, which cannot define a holographic map for the baby universe due to its lack of an asymptotic boundary. Therefore, we consider an alternative perspective: there are actually \textit{two} different holographic maps, one for each candidate bulk state.

As a result, the boundary operator $\swap_\partial$ admits two consistent bulk duals -- we will call them $\swap$ and $\swap_{\baby}$ -- depending on the choice of holographic map. We find that measuring $\swap_\partial$ on the boundary cannot distinguish between measuring $\swap$ in the no-baby state and measuring $\swap_{\baby}$ in the baby state. This interpretation of EG's swap test is more closely aligned with AR's ``beyond AdS/CFT'' hypothesis, reflecting the non-uniqueness of the bulk reconstruction rather than a failure of the semiclassical description.

The remainder of this work will be organized as follows. In Section \ref{sec:second_map}, we define the two holographic maps for both of AR's bulk duals and identify the respective bulk operators dual to $\swap_\partial$. We will construct a toy qubit model to further illustrate the properties of these holographic maps and bulk operators. Section \ref{sec:simplicity} examines the simplicity of $\swap_\partial$ and its implications for semiclassical validity. Section \ref{sec:obs} considers how the rules of \cite{harlow_Quantum_2025,akers_Observers_2025} might be applied to include observers in the baby universe, and section \ref{sec:conc} concludes with a few final remarks.

\textit{Note to the reader:} during preparation of this work, the author learned of related results independently prepared in \cite{engelhardt_observer_2025}, to appear on the arXiv simultaneously.

\section{A tale of two holographic maps} \label{sec:second_map}

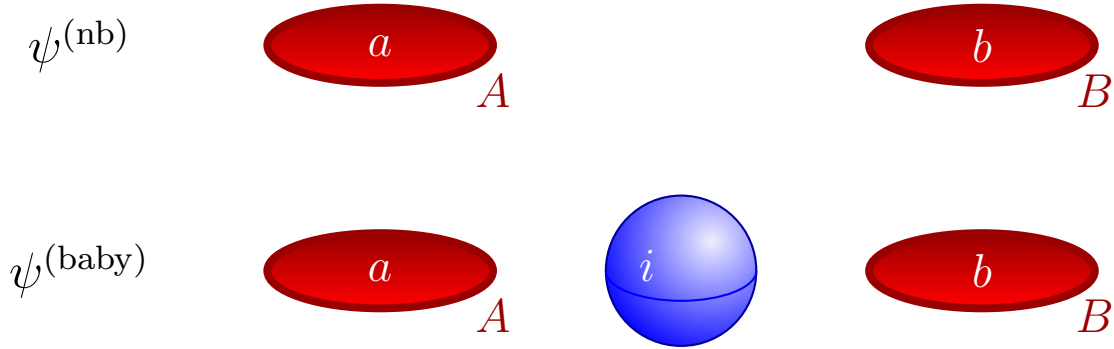
\begin{figure}
    \centering
    \input{baby_no_baby}
    \caption{A graphical representation of the no-baby state $\psi^{(\nb)}$ (top) and baby state $\psi^{(\baby)}$ (bottom). The red discs labeled ($a$, $b$) represent AdS spacetimes with asymptotic boundaries ($A$, $B$). The blue ball labeled $i$ represents a baby (or \textit{closed}) universe.}
    \label{fig:baby_no_baby}
\end{figure}

We construct two holographic maps for the candidate bulk states found by AR, each consistent with the extrapolate dictionary. We begin by recalling that both bulk states contain two disconnected AdS spacetimes, while only one includes a baby universe. We will denote these states in this work as follows:
\begin{align}
    \text{``No-baby'' (nb) state:}&\qquad \psi^{(\nb)} \in \hs_{\nb} = \hs_a \otimes \hs_b \\
    \text{``Baby'' state:}&\qquad \psi^{(\baby)} \in \hs_{\baby} = \hs_a \otimes \hs_b \otimes \hs_i \\
    \text{CFT state:}&\qquad \Psi^{(\partial)} \in \hs_\partial = \hs_A \otimes \hs_B,
\end{align}
where ($a$, $b$) denotes (left, right) disconnected AdS spacetimes, ($A$, $B$) denotes their respective boundary CFTs, and $i$ denotes the baby universe; see figure \ref{fig:baby_no_baby} for a graphical representation of these states. Note that both AR and EG referred to the baby state as ``description 1'' and the no-baby state as ``description 2''; we have altered the notation here to make it easier to distinguish the states and their respective holographic maps.

We define the first holographic map as one that maps the no-baby bulk state to the boundary CFT state,
\begin{equation}
    V_{\nb} : \hs_{\nb} \to \hs_\partial, \qquad \Psi^{(\partial)} = V_{\nb} \psi^{(\nb)} V_{\nb}^\dagger.
\end{equation}
Just as in \cite{engelhardt_Further_2025}, we take $V_{\nb}$ to be defined solely by the extrapolate dictionary. As a result, the no-baby holographic map is isometric, $V_{\nb}^\dagger V_{\nb} = \id_{ab}$. Similarly, the second holographic map will act as a map from the baby state to the same boundary CFT state,
\begin{equation} \label{eq:V_baby_gen}
    V_{\baby} : \hs_{\baby} \to \hs_\partial, \qquad \Psi^{(\partial)} = V_{\baby} \psi^{(\baby)} V_{\baby}^\dagger.
\end{equation}
We define this new holographic map to act as
\begin{equation}
    V_{\baby} \equiv |i|^{1/2} V_{\nb} \otimes \langle\chi|_i,
\end{equation}
where $\langle\chi|_i$ is post-selection on the entire baby universe $\hs_i\subset\hs_{\baby}$, making $V_{\baby}$ non-isometric. At this point there is no special choice of $\langle\chi|_i$, so we take $\langle\chi|_i$ to be random and include the prefactor $|i| \equiv \dim \hs_i$ to preserve state normalization on average. Note that we have taken $V_{\baby}$ to act as the same isometry $V_{\nb}$ on the $\hs_a\otimes\hs_b$ subsystem of $\hs_{\baby}$, since the extrapolate dictionary implies that both holographic maps should act identically on the disconnected AdS components. 

Consistency with AR's results \cite{antonini_Holographic_2025} requires that both bulk states be mapped to the same boundary state $\Psi^{(\partial)}$ by their respective holographic maps:
\begin{equation} \label{eq:nb_baby_rel}
    V_{\nb} \psi^{(\nb)} V_{\nb}^\dagger \approx V_{\baby} \psi^{(\baby)} V^\dagger_{\baby}.
\end{equation}
Using the fact that $V_{\baby}$ acts as the isometry $V_{\nb}$ on the $\hs_a \otimes \hs_b$ subsystem, we find that this implies the following relationship between the two bulk states:
\begin{equation} \label{eq:nb_baby_rel2}
    \psi^{(\nb)} \approx |i| \, (\id_{ab} \otimes \langle\chi|_i ) \psi^{(\baby)} (\id_{ab} \otimes |\chi\rangle_i ).
\end{equation}
In order for both bulk states to be dual to the same boundary state, the bulk states cannot be independent, but are related by the state $|\chi\rangle$.

As maps on Hilbert spaces, both $V_{\nb}$ and $V_{\baby}$ induce operator maps from the boundary back to their respective bulks. This can be seen from the cyclicity of the trace; given a generic holographic map $V:\hs_{\text{bulk}} \to \hs_\partial$ and a boundary operator $O_\partial \in \B(\hs_\partial)$, its expectation value
\begin{align*}
    \langle O_\partial \rangle_{\Psi^{(\partial)}} &= \tr \big[ \Psi^{(\partial)} O_\partial \big] \\
        &= \tr \big[ V \psi^{(\text{bulk})} V^\dagger O_\partial \big] \\
        &= \tr \big[ \psi^{(\text{bulk})} V^\dagger O_\partial V \big] \\
        &= \langle O_\text{bulk} \rangle_{\psi^{(\text{bulk})}} \numberthis \label{eq:opMap}
\end{align*}
is equivalent to the expectation value of a bulk operator $O_\text{bulk} \in \B(\hs_\text{bulk})$ given by
\begin{equation} \label{eq:opMap2}
    O_\text{bulk} \equiv V^\dagger O_\partial V.
\end{equation}
In this way, the operator map induced by $V_{\nb}$, denoted $V_{\nb}^* : \B(\hs_\partial)\to\B(\hs_{\nb})$, maps the boundary operator $\swap_\partial$ to its bulk dual $\swap$,
\begin{equation} \label{eq:SbdyToS}
    \swap = (V_{\nb}^\dagger \otimes V_{\nb}^\dagger) \swap_\partial (V_{\nb} \otimes V_{\nb}).
\end{equation}
This is the same bulk swaperator $\swap\in \B(\hs_{\nb}\otimes\hs_{\nb'})$ defined by EG to swap the AdS spacetimes ($a$, $b$) with copies ($a'$, $b'$),
\begin{equation}
    \swap \big( |\psi\rangle_{ab} \otimes |\phi\rangle_{a'b'} \big) = |\phi\rangle_{ab} \otimes |\psi\rangle_{a'b'}.
\end{equation}
Thanks to the isometry of $V_{\nb}$, (\ref{eq:SbdyToS}) is consistent with the reconstruction criteria $\swap_\partial (V_{\nb} \otimes V_{\nb}) = (V_{\nb} \otimes V_{\nb}) \swap$. 

At no point did we require $V$ to be isometric in (\ref{eq:opMap}); the operator map (\ref{eq:opMap2}) is just as valid for non-isometric maps. Therefore, $V_{\baby}$ induces a second operator map, denoted $V^*_{\baby}:\B(\hs_\partial)\to\B(\hs_{\baby})$, taking the same boundary operator $\swap_\partial$ to
\begin{equation} \label{eq:def_Sbaby}
    \swap_{\baby} \equiv (V^\dagger_{\baby} \otimes V^\dagger_{\baby}) \swap_\partial (V_{\baby} \otimes V_{\baby}) = |i| \, \swap \otimes |\chi\rangle\langle\chi|_i \otimes |\chi\rangle\langle\chi|_{i'},
\end{equation}
where $i$ and $i'$ are two copies of the baby universe. Thus the bulk operator dual to $\swap_\partial$ in the baby state is not the bulk swaperator $\swap$! Instead, it is a projector acting on the two baby universe copies in addition to the swaperator acting on the disconnected AdS spacetimes. Following (\ref{eq:opMap}) again, let us check that the expectation value of $\swap_{\baby}$ in the baby state matches that of $\swap_\partial$,
\begin{align*}
    \langle \swap_\partial \rangle_{\Psi^{(\partial)}} &= \tr \big[ ( \Psi^{(\partial)} \otimes \Psi^{(\partial)} ) \swap_\partial \big] \\
        &= \tr \big[ ( V_{\baby} \otimes V_{\baby} ) ( \psi^{(\baby)} \otimes \psi^{(\baby)} ) ( V^\dagger_{\baby} \otimes V^\dagger_{\baby} ) \swap_\partial \big] \\
        &= \tr \big[ ( \psi^{(\baby)} \otimes \psi^{(\baby)} ) ( V^\dagger_{\baby} \otimes V^\dagger_{\baby} ) \swap_\partial ( V_{\baby} \otimes V_{\baby} ) \big] \\
        &= \tr \big[ ( \psi^{(\baby)} \otimes \psi^{(\baby)} ) \swap_{\baby} \big] \\
        & = \langle\swap_{\baby}\rangle_{\psi^{(\baby)}}, \numberthis
\end{align*}
where we used (\ref{eq:V_baby_gen}) in the second line, the cyclic property of the trace in the third, and (\ref{eq:def_Sbaby}) in the fourth. We now have different bulk operators for the two bulk states, both of which yield the same expectation value as $\swap_\partial$:
\begin{equation} \label{eq:tri_equal}
    \langle\swap_\partial\rangle_{\Psi^{(\partial)}} = \langle\swap\rangle_{\psi^{(\nb)}} = \langle\swap_{\baby}\rangle_{\psi^{(\baby)}}.
\end{equation}
Measuring $\swap_\partial$ on the boundary therefore does not correspond to measuring $\swap$ in the baby state, but rather to measuring $\swap_{\baby}$. This indicates that the boundary operator $\swap_\partial$ cannot distinguish between the two states, since this measurement can be equally well described by measurements of two different dual bulk operators on the two candidate bulk states.

\subsection{Qubit model}
To further illustrate the properties of $V_{\baby}$ and $\swap_{\baby}$, we present a toy example. We model the entanglement structure of the baby universe state by two pairs of Bell states $|\Phi^+\rangle$ on qubits:
\begin{equation} \label{eq:ex_baby}
    |\psi^{(\baby)}\rangle = |\Phi^+\rangle_{a,i_1} |\Phi^+\rangle_{b,i_2},
\end{equation}
where we have split the baby universe into $\hs_i = \hs_{i_1} \otimes \hs_{i_2}$. For further simplicity, we take the extrapolate dictionary to be trivial such that
\begin{equation} \label{eq:ex_V}
    V_{\nb} = \id_{ab}, \qquad V_{\baby} = 2 \, \id_{ab} \otimes \langle\chi|_{i_1 i_2}.
\end{equation}
The dual CFT state is then given by
\begin{equation}
    |\Psi^{(\partial)}\rangle = V_{\baby} |\psi^{(\baby)}\rangle = |\chi^*\rangle_{ab},
\end{equation}
and (\ref{eq:nb_baby_rel}) requires the no-baby state be given by
\begin{equation}
    |\psi^{(\nb)}\rangle = V_{\nb}^\dagger V_{\baby} |\psi^{(\baby)}\rangle = |\chi^*\rangle_{ab}.
\end{equation}
The equality of $|\psi^{(\nb)}\rangle$ and $|\Psi^{(\partial)}\rangle$ follows from the triviality of $V_{\nb}$. We note that the dependence on the choice of post-selection $\langle\chi|_i$ becomes very explicit in this example: the post-selection must be chosen such that $|\chi^*\rangle_{ab}$ matches the desired no-baby state. For example, we might choose $\langle\chi|_i$ to be the maximally entangled state $\langle\Phi^+|_{i_1, i_2}$ if we want the no-baby state to be entangled between the disconnected AdS spacetimes $a$ and $b$. Since our primary results do not depend on this choice, we will leave $\langle\chi|_i$ generic.

Because $|\chi^*\rangle_{ab}$ is pure, the expectation values of both $\swap$ and $\swap_\partial$ are 1, as was found in \cite{engelhardt_Further_2025}. We now confirm that the same is true of $\swap_{\baby}$ in the baby state:
\begin{align*}
    \langle \swap_{\baby} \rangle &= 16 \big( \langle\Phi^+|_{a,i_1} \langle\Phi^+|_{b,i_2} \langle\Phi^+|_{a',i'_1} \langle\Phi^+|_{b',i'_2} \big) \\
        & \qquad \times \big( \swap \otimes |\chi\rangle\langle\chi|_{i_1 i_2} \otimes |\chi\rangle\langle\chi|_{i'_1 i'_2} \big) \, \big( |\Phi^+\rangle_{a,i_1} |\Phi^+\rangle_{b,i_2} |\Phi^+\rangle_{a',i'_1} |\Phi^+\rangle_{b',i'_2} \big) \\
        &= \big( \langle\chi^*|_{ab} \langle\chi^*|_{a'b'} \big) \swap \big( |\chi^*\rangle_{ab} |\chi^*\rangle_{a'b'} \big) \\
        &= 1. \numberthis
\end{align*}
Had we instead evaluated the expectation value of the swaperator $\swap \otimes \id_i \otimes \id_{i'}$ in the baby state, we would have found $1/4$, consistent with equation (\ref{eq:exp_vals}). Because of the inclusion of projectors $|\chi\rangle\langle\chi|_i$ on the baby universe, $\swap_{\baby}$ has the same expectation value as both bulk operators $\swap$ and $\swap_\partial$ in their respective states. 

Given the second holographic map $V_{\baby}$, we have shown that $\swap_{\baby}$ -- not the original swaperator $\swap$ -- is the correct bulk dual of $\swap_\partial$ in the baby state. We therefore interpret $\swap_{\baby}$ as the true, non-perturbatively correct version of the ``naive'' bulk operator $\swap$. So long as we use the correct non-perturbative answer, the swap test will not be able to distinguish between AR's two candidate bulk states.

\section{Simplicity of $\swap_\partial$} \label{sec:simplicity}

An emerging principle in holography is that operator complexity constrains the validity of semiclassical gravity as an effective description of bulk physics 
\cite{harlow_Quantum_2013,
susskind_Computational_2016,
engelhardt_Decoding_2018,
engelhardt_Coarse_2019,
brown_Pythons_2020,
bouland_Computational_2019,
kim_Ghost_2020,
engelhardt_World_2021,
engelhardt_Finding_2022,
akers_Black_2024}. 
This idea partly motivated EG's use of the swap test, as the boundary operator $\swap_\partial$ is \textit{simple}. When a simple operator can distinguish naive perturbative predictions from correct non-perturbative results -- such as after a black hole has fully evaporated -- it suggests that the perturbative description no longer accurately captures the bulk physics. 

This appears to be the case here: the simple operator $\swap_\partial$ agrees with the non-perturbative result $\langle\swap_{\baby}\rangle_{\psi^{(\baby)}}$ (by construction) but is easily distinguishable from the naive perturbative prediction $\langle\swap\rangle_{\psi^{(\baby)}}$. Therefore, even though we have found the non-perturbative operator $\swap_{\baby}$ that provides the correct answer, we might still suspect that $\psi^{(\baby)}$ is not a valid semiclassical dual for $\Psi^{(\partial)}$.

However, we demonstrate here that $\swap_\partial$ is not able to distinguish perturbative and non-perturbative answers in the baby state universally. Instead, there exists a regime in which both answers are exponentially small, making them indistinguishable by a simple operator. The existence of such a regime suggests we should not hastily rule out $\psi^{(\baby)}$ as a viable semiclassical dual.\footnote{We thank Chris Akers for discussions on this point.}

To demonstrate this, let us introduce an external reference $R$ that we will entangle with our boundary CFT state. Consider AR's preparation of $|\Psi^{(\partial)}\rangle_{AB}$ via the insertion of a heavy operator $O$ in the Euclidean evolution of the two boundary CFTs \cite{antonini_Holographic_2025}. We can entangle the CFT state on $AB$ with $R$ by adding flavor degrees of freedom (indexed by $k$) to $O$:
\begin{equation} \label{eq:Ok}
    |\Psi^{(\partial)}\rangle_{ABR} = \frac{1}{\sqrt{Z|R|}} \sum_{k,m,n} \left( e^{-(\beta_A E_m + \beta_B E_n)/2} O^{(k)}_{m,n} |E_m\rangle_A|E_n\rangle_B \right) |k\rangle_R,
\end{equation}
where $|R|\equiv\dim\hs_R$ and $k$ runs over $\{1,\dots,|R|\}$. The entanglement structures of both bulk states are modified by this new entanglement. The no-baby state is now purified by the reference $R$, so that $S(ab) = S(R)$. The corresponding expectation value of $\swap$ in the no-baby state is
\begin{equation}
    \langle\swap\rangle_{\psi^{(\nb)}} = e^{-S_2(R)},
\end{equation}
exponentially suppressed by the Renyi entropy of $R$. By (\ref{eq:tri_equal}), we know the same must be true of the non-perturbative expectation value $\langle\swap_{\baby}\rangle_{\psi^{(\baby)}}$ in the baby state.

Alternatively, the two AdS components $ab$ of the baby state are now purified by both the baby universe $i$ and reference $R$. However, the heaviness of the inserted operator $O^{(k)}$ causes the corresponding bulk excitation (or ``dust ball'') to fall into the baby universe. Therefore the reference system is purified \textit{only} by the baby universe $i$, and no entanglement exists between $ab$ and $R$ in the baby state. The expectation value of $\swap$ in the baby state is independent of the entanglement with $R$,
\begin{equation}
    \langle\swap\rangle_{\psi^{(\baby)}} = e^{-S_2(ab)} = e^{-S_2(iR)},
\end{equation}
and therefore unchanged by the addition of the reference. The naive perturbative prediction $\langle\swap\rangle_{\psi^{(\baby)}}$ remains exponentially small in the Renyi entropy of $ab$.

Therefore, when the boundary state $\Psi^{(\partial)}$ is entangled with a reference system $R$ in this way, both the naive perturbative prediction $\langle\swap\rangle_{\psi^{(\baby)}}$ and the non-perturbative result $\langle\swap_{\baby}\rangle_{\psi^{(\baby)}}$ become exponentially small. In this regime, the simple boundary operator $\swap_\partial$ is unable to distinguish between the two, since both outcomes are similarly suppressed. This demonstrates that there exist regimes in which $\psi^{(\baby)}$ cannot be ruled out by the simplicity of $\swap_\partial$, preserving its viability as a semiclassical description of the baby universe.

\section{Including observers} \label{sec:obs}

After showing that $\swap_{\baby}$ evades the swap test and arguing that there are regimes where the semiclassical description of the baby universe is valid, it might still seem that little is gained by defining $V_{\baby}$. After all, the baby universe's fundamental Hilbert space remains one-dimensional, and the only bulk operators naturally dual to $\B(\hs_\partial)$ are those involving projectors on $\hs_i$. Furthermore, we already had a perfectly well-defined bulk dual in $\psi^{(\nb)}$. Why sacrifice the uniqueness of the holographic map just to preserve the validity of a redundant description? Nevertheless, defining $V_{\baby}$ in this way provides one key advantage: it enables us to study observers in the baby universe using recently developed techniques. 

There are currently two proposals for incorporating observers into holographic maps. The first -- proposed by Harlow, Usatyuk, and Zhao (HUZ) \cite{harlow_Quantum_2025} -- takes the observer to be classical and clones them into an external reference in their pointer basis. The second -- proposed by Akers, Bueller, DeWolfe, Higginbotham, Reinking, and Rodriguez (the ``Colorado group'') \cite{akers_Observers_2025} and inspired by \cite{abdalla_Gravitational_2025} -- assumes the observer is already in the fundamental description and removes the part of the holographic map acting on their local patch.\footnote{The holographic code prescription used by the Colorado group was also recently used in \cite{kaya_Hollowgrams_2025} to define ``hollowed'' random tensor networks implementing Bousso and Penington's generalized entanglement wedge prescription \cite{bousso_Entanglement_2023,bousso_Holograms_2023}.}

Either approach can be readily applied to $V_{\baby}$ by interpreting the heavy operator $O$ in (\ref{eq:Ok}) as the insertion of some observer. Using either the HUZ or Colorado rules, the resulting modification of $V_{\baby}$ would lead to an enlargement of the fundamental Hilbert space by an observer-dependent factor $\hs_I$,
\begin{equation}
    \hs_\text{fun} = \hs_A \otimes \hs_B \otimes \hs_I.
\end{equation}
This yields a non-trivial Hilbert space for the baby universe, allowing semiclassical states in $\hs_i$ to be encoded holographically. Furthermore, new operators on $\B(\hs_I)$ will be naturally mapped by the modified operator map $V^*_{\baby}$ to non-trivial operators on $\hs_i$. We note that while both approaches will produce these non-trivial tensor factors and operators for the baby universe, they will differ in the resulting dimension of $\hs_I$ and the form of bulk operators in $\B(\hs_i)$.

Note that in either case, the observer can serve as the reference used in section \ref{sec:simplicity} to demonstrate validity of the semiclassical description given the simplicity of $\swap_\partial$. This is quite natural from the perspective of the HUZ rules, where the external clone is very reminiscent of the entangled external reference. The Colorado rules are no different; by removing the portion of the map acting on the observer, the observer is effectively treated as an external reference from the perspective of the fundamental description. So long as they are entangled with their surrounding environment within the semiclassical closed universe (a reasonable assumption) the Colorado observer can also provide the external entanglement necessary for section \ref{sec:simplicity}.

\section{Concluding remarks} \label{sec:conc}

We have argued that allowing for post-selection on baby universes is consistent with the extrapolate dictionary, which permits the definition of two holographic maps, $V_{\nb}$ and $V_{\baby}$, for AR's two candidate bulk states $\psi^{(\nb)}$ and $\psi^{(\baby)}$. These maps naturally yield two distinct bulk operators, $\swap$ and $\swap_{\baby}$, implementing the swap test in their respective bulk states. Since both bulk operators have the same expectation value as their common boundary dual $\swap_\partial$, a measurement of $\swap_\partial$ cannot distinguish the candidate bulk states. 

Moreover, the simplicity of $\swap_\partial$ does not invalidate a semiclassical baby universe in all regimes. When the baby universe is entangled with an external reference $R$, the naive measurement $\langle\swap\rangle_{\psi^{(\baby)}}$ and non-perturbative result $\langle\swap_{\baby}\rangle_{\psi^{(\baby)}}$ become indistinguishable to sub-exponentially complex experiments. This entanglement could be provided by an observer, which can now be included in the baby universe by applying the rules of \cite{harlow_Quantum_2025,akers_Observers_2025} to modify $V_{\baby}$. 

Finally, we might wonder whether the observers discussed in section \ref{sec:obs} could provide the extra ``beyond AdS/CFT'' data needed to distinguish the two bulk states and resolve the AR puzzle. If an observer in the baby universe can be equivalently described by some other observer in $\hs_{\nb}$, then the puzzle remains. For this to be the case, a version of equation (\ref{eq:nb_baby_rel}) would need to hold for both $V_{\baby}$ and $V_{\nb}$ modified by their respective observers. If no such equivalent observer exists in the no-baby state, then the puzzle appears to be resolved. We leave these intriguing questions for future work.

\section*{Acknowledgements}

I am grateful to Chris Akers, Gracemarie Bueller, and Oliver DeWolfe for helpful discussions and comments on the draft. This work was supported by the Department of Energy under grant DE-SC0010005.

\bibliographystyle{JHEP}
\bibliography{prego}

\end{document}

%% file: baby_no_baby.tex
\begin{tikzpicture}

\newcommand{\AdS}[3]{
    \shade[bottom color=red, top color=red!60!black] 
        (#1,#2) ellipse [x radius=1.5cm, y radius=0.5cm];
    \draw[red!60!black, line width=3pt] 
        (#1,#2) ellipse [x radius=1.5cm, y radius=0.5cm];
    \node[scale=1.5,white] at (#1,#2) {$#3$};
    \node[scale=1.5,red!60!black] at (#1,#2) [shift={(1,-0.4)}] {$\uppercase{#3}$};
}

\newcommand{\babyPic}[2]{
    \def\ballcenter{(#1,#2)};
    \def\radius{1cm}; 
    \begin{scope}
        \clip \ballcenter circle [radius=\radius];
        \shade[inner color=blue!7!white, outer color=blue] 
            \ballcenter + (0.4,0.4) circle [radius=1.6cm]; 
    \end{scope}
    \draw[blue!60!black, thick] \ballcenter circle [radius=\radius];
    \draw[thick, blue]
        \ballcenter + (\radius,0)
        [xscale=1, yscale=0.4]
        arc[start angle=0, end angle=-180, radius=1];
    \node[scale=1.5,white] at \ballcenter [shift={(-0.3,0.05)}] {$i$};
}

\node[scale=1.6] at (-8,0) {$\psi^{(\baby)}$};
\AdS{-4}{0}{a};
\babyPic{0}{0};
\AdS{4}{0}{b};

\node[scale=1.6] at (-8,3) {$\psi^{(\nb)}$};
\AdS{-4}{3}{a};
\AdS{4}{3}{b};

\end{tikzpicture}

%% file: main.bbl
\providecommand{\href}[2]{#2}\begingroup\raggedright\begin{thebibliography}{10}

\bibitem{maldacena_Wormholes_2004}
J.~Maldacena and L.~Maoz, \emph{Wormholes in {{AdS}}}, \href{https://doi.org/10.1088/1126-6708/2004/02/053}{\emph{Journal of High Energy Physics} {\bfseries 2004} (2004) 053}.

\bibitem{almheiri_Page_2020}
A.~Almheiri, R.~Mahajan, J.~Maldacena and Y.~Zhao, \emph{The {{Page}} curve of {{Hawking}} radiation from semiclassical geometry}, \href{https://doi.org/10.1007/JHEP03(2020)149}{\emph{Journal of High Energy Physics} {\bfseries 2020} (2020) 149} [\href{https://arxiv.org/abs/1908.10996}{{\ttfamily 1908.10996}}].

\bibitem{penington_Replica_2020}
G.~Penington, S.H.~Shenker, D.~Stanford and Z.~Yang, \emph{Replica wormholes and the black hole interior},  \href{https://arxiv.org/abs/1911.11977}{{\ttfamily 1911.11977}}.

\bibitem{marolf_Transcending_2020}
D.~Marolf and H.~Maxfield, \emph{Transcending the ensemble: Baby universes, spacetime wormholes, and the order and disorder of black hole information}, \href{https://doi.org/10.1007/JHEP08(2020)044}{\emph{Journal of High Energy Physics} {\bfseries 2020} (2020) 44}.

\bibitem{mcnamara_Baby_2020}
J.~McNamara and C.~Vafa, \emph{Baby {{Universes}}, {{Holography}}, and the {{Swampland}}},  \href{https://arxiv.org/abs/2004.06738}{{\ttfamily 2004.06738}}.

\bibitem{usatyuk_Closed_2024}
M.~Usatyuk, Z.-Y.~Wang and Y.~Zhao, \emph{Closed universes in two dimensional gravity}, \href{https://doi.org/10.21468/SciPostPhys.17.2.051}{\emph{SciPost Physics} {\bfseries 17} (2024) 051}.

\bibitem{usatyuk_Closed_2025}
M.~Usatyuk and Y.~Zhao, \emph{Closed universes, factorization, and ensemble averaging}, \href{https://doi.org/10.1007/JHEP02(2025)052}{\emph{Journal of High Energy Physics} {\bfseries 2025} (2025) 52}.

\bibitem{harlow_Quantum_2025}
D.~Harlow, M.~Usatyuk and Y.~Zhao, \emph{Quantum mechanics and observers for gravity in a closed universe},  \href{https://arxiv.org/abs/2501.02359}{{\ttfamily 2501.02359}}.

\bibitem{abdalla_Gravitational_2025}
A.I.~Abdalla, S.~Antonini, L.V.~Iliesiu and A.~Levine, \emph{The gravitational path integral from an observer's point of view},  \href{https://arxiv.org/abs/2501.02632}{{\ttfamily 2501.02632}}.

\bibitem{akers_Observers_2025}
C.~Akers, G.~Bueller, O.~DeWolfe, K.~Higginbotham, J.~Reinking and R.~Rodriguez, \emph{On observers in holographic maps}, \href{https://doi.org/10.1007/JHEP05(2025)201}{\emph{Journal of High Energy Physics} {\bfseries 2025} (2025) 201}.

\bibitem{chen_Observers_2025}
H.Z.~Chen, \emph{Observers seeing gravitational {{Hilbert}} spaces: Abstract sources for an abstract path integral},  \href{https://arxiv.org/abs/2505.15892}{{\ttfamily 2505.15892}}.

\bibitem{antonini_Holographic_2025}
S.~Antonini and P.~Rath, \emph{Do holographic {{CFT}} states have unique semiclassical bulk duals?},  \href{https://arxiv.org/abs/2408.02720}{{\ttfamily 2408.02720}}.

\bibitem{maldacena_Large_1999}
J.M.~Maldacena, \emph{The {{Large N Limit}} of {{Superconformal Field Theories}} and {{Supergravity}}}, \href{https://doi.org/10.1023/A:1026654312961}{\emph{International Journal of Theoretical Physics} {\bfseries 38} (1999) 1113} [\href{https://arxiv.org/abs/hep-th/9711200}{{\ttfamily hep-th/9711200}}].

\bibitem{gubser_Gauge_1998}
S.S.~Gubser, I.R.~Klebanov and A.M.~Polyakov, \emph{Gauge {{Theory Correlators}} from {{Non-Critical String Theory}}}, \href{https://doi.org/10.1016/S0370-2693(98)00377-3}{\emph{Physics Letters B} {\bfseries 428} (1998) 105} [\href{https://arxiv.org/abs/hep-th/9802109}{{\ttfamily hep-th/9802109}}].

\bibitem{witten_Sitter_1998}
E.~Witten, \emph{Anti de {{Sitter}} space and holography}, \href{https://doi.org/10.4310/ATMP.1998.v2.n2.a2}{\emph{Advances in Theoretical and Mathematical Physics} {\bfseries 2} (1998) 253}.

\bibitem{jackiw_Lower_1985}
R.~Jackiw, \emph{Lower dimensional gravity}, \href{https://doi.org/10.1016/0550-3213(85)90448-1}{\emph{Nuclear Physics B} {\bfseries 252} (1985) 343}.

\bibitem{teitelboim_Gravitation_1983}
C.~Teitelboim, \emph{Gravitation and hamiltonian structure in two spacetime dimensions}, \href{https://doi.org/10.1016/0370-2693(83)90012-6}{\emph{Physics Letters B} {\bfseries 126} (1983) 41}.

\bibitem{saad_JT_2019}
P.~Saad, S.H.~Shenker and D.~Stanford, \emph{{{JT}} gravity as a matrix integral},  \href{https://arxiv.org/abs/1903.11115}{{\ttfamily 1903.11115}}.

\bibitem{engelhardt_Further_2025}
N.~Engelhardt and E.~Gesteau, \emph{Further {{Evidence Against}} a {{Semiclassical Baby Universe}} in {{AdS}}/{{CFT}}},  \href{https://arxiv.org/abs/2504.14586}{{\ttfamily 2504.14586}}.

\bibitem{antonini_Cosmology_2023}
S.~Antonini, M.~Sasieta and B.~Swingle, \emph{Cosmology from random entanglement},  \href{https://arxiv.org/abs/2307.14416}{{\ttfamily 2307.14416}}.

\bibitem{akers_Black_2024}
C.~Akers, N.~Engelhardt, D.~Harlow, G.~Penington and S.~Vardhan, \emph{The black hole interior from non-isometric codes and complexity}, \href{https://doi.org/10.1007/JHEP06(2024)155}{\emph{Journal of High Energy Physics} {\bfseries 2024} (2024) 155}.

\bibitem{engelhardt_observer_2025}
N.~Engelhardt, E.~Gesteau and D.~Harlow, \emph{Observer complementarity for black holes and holography},  \href{https://arxiv.org/abs/2507.06046}{{\ttfamily 2507.06046}}.

\bibitem{harlow_Quantum_2013}
D.~Harlow and P.~Hayden, \emph{Quantum {{Computation}} vs. {{Firewalls}}}, \href{https://doi.org/10.1007/JHEP06(2013)085}{\emph{Journal of High Energy Physics} {\bfseries 2013} (2013) 85} [\href{https://arxiv.org/abs/1301.4504}{{\ttfamily 1301.4504}}].

\bibitem{susskind_Computational_2016}
L.~Susskind, \emph{Computational complexity and black hole horizons}, \href{https://doi.org/10.1002/prop.201500092}{\emph{Fortschritte der Physik} {\bfseries 64} (2016) 24}.

\bibitem{engelhardt_Decoding_2018}
N.~Engelhardt and A.C.~Wall, \emph{Decoding the {{Apparent Horizon}}: {{Coarse-Grained Holographic Entropy}}}, \href{https://doi.org/10.1103/PhysRevLett.121.211301}{\emph{Physical Review Letters} {\bfseries 121} (2018) 211301}.

\bibitem{engelhardt_Coarse_2019}
N.~Engelhardt and A.C.~Wall, \emph{Coarse graining holographic black holes}, \href{https://doi.org/10.1007/JHEP05(2019)160}{\emph{Journal of High Energy Physics} {\bfseries 2019} (2019) 160}.

\bibitem{brown_Pythons_2020}
A.R.~Brown, H.~Gharibyan, G.~Penington and L.~Susskind, \emph{The {{Python}}'s {{Lunch}}: Geometric obstructions to decoding {{Hawking}} radiation}, \href{https://doi.org/10.1007/JHEP08(2020)121}{\emph{Journal of High Energy Physics} {\bfseries 2020} (2020) 121}.

\bibitem{bouland_Computational_2019}
A.~Bouland, B.~Fefferman and U.~Vazirani, \emph{Computational pseudorandomness, the wormhole growth paradox, and constraints on the {{AdS}}/{{CFT}} duality},  \href{https://arxiv.org/abs/1910.14646}{{\ttfamily 1910.14646}}.

\bibitem{kim_Ghost_2020}
I.H.~Kim, E.~Tang and J.~Preskill, \emph{The ghost in the radiation: {{Robust}} encodings of the black hole interior}, \href{https://doi.org/10.1007/JHEP06(2020)031}{\emph{Journal of High Energy Physics} {\bfseries 2020} (2020) 31} [\href{https://arxiv.org/abs/2003.05451}{{\ttfamily 2003.05451}}].

\bibitem{engelhardt_World_2021}
N.~Engelhardt, G.~Penington and A.~{Shahbazi-Moghaddam}, \emph{A {{World}} without {{Pythons}} would be so {{Simple}}}, \href{https://doi.org/10.1088/1361-6382/ac2de5}{\emph{Classical and Quantum Gravity} {\bfseries 38} (2021) 234001} [\href{https://arxiv.org/abs/2102.07774}{{\ttfamily 2102.07774}}].

\bibitem{engelhardt_Finding_2022}
N.~Engelhardt, G.~Penington and A.~{Shahbazi-Moghaddam}, \emph{Finding {{Pythons}} in {{Unexpected Places}}}, \href{https://doi.org/10.1088/1361-6382/ac3e75}{\emph{Classical and Quantum Gravity} {\bfseries 39} (2022) 094002} [\href{https://arxiv.org/abs/2105.09316}{{\ttfamily 2105.09316}}].

\bibitem{kaya_Hollowgrams_2025}
S.~Kaya, P.~Rath and K.~Ritchie, \emph{Hollow-grams: {{Generalized Entanglement Wedges}} from the {{Gravitational Path Integral}}},  \href{https://arxiv.org/abs/2506.10064}{{\ttfamily 2506.10064}}.

\bibitem{bousso_Entanglement_2023}
R.~Bousso and G.~Penington, \emph{Entanglement wedges for gravitating regions}, \href{https://doi.org/10.1103/PhysRevD.107.086002}{\emph{Physical Review D} {\bfseries 107} (2023) 086002}.

\bibitem{bousso_Holograms_2023}
R.~Bousso and G.~Penington, \emph{Holograms in our world}, \href{https://doi.org/10.1103/PhysRevD.108.046007}{\emph{Physical Review D} {\bfseries 108} (2023) 046007}.

\end{thebibliography}\endgroup
